\documentclass[]{vgtc}            
\ifpdf
  \pdfoutput=1\relax                   
  \usepackage{graphicx}                
  \DeclareGraphicsExtensions{.pdf,.png,.jpg,.jpeg} 
\else
  \usepackage{graphicx}                
  \DeclareGraphicsExtensions{.eps}     
\fi%

\graphicspath{{figures/}{pictures/}{images/}{./}} 

\usepackage{microtype}                 
\PassOptionsToPackage{warn}{textcomp}  
\usepackage{textcomp}                  
\usepackage{mathptmx}                  
\usepackage{times}                     
\usepackage{cite}                      
\usepackage{tabu}                      
\usepackage{booktabs}                  

\usepackage{subcaption}
\usepackage{acronym}
\usepackage{hyperref}
\usepackage{amsmath}

\usepackage{url}
\usepackage{todonotes}
\let\xtodo\todo
\renewcommand{\todo}[1]{\xtodo[inline,color=green!50]{#1}}

\onlineid{8836}

\vgtccategory{Human-Computer Interaction, Virtual Reality, Perception}

\preprinttext{To appear at IEEE ISMAR.}

\newcommand{\added}[1]{\textcolor{black}{#1}}

\title{Mind the Visual Discomfort: Assessing Event-Related Potentials as Indicators for Visual Strain in Head-Mounted Displays}

\author{Francesco Chiossi\thanks{e-mail: francesco.chiossi@um.ifi.lmu.de}\\ %
        \scriptsize LMU Munich, Munich %
\and Yannick Weiss\thanks{e-mail: yannick.weiss@um.ifi.lmu.de}\\ %
     \scriptsize LMU Munich, Munich %
\and Thomas Steinbrecher\thanks{e-mail: thomas.steinbrecher@campus.lmu.de}\\ %
     \scriptsize LMU Munich, Munich %
\and Christian Mai\thanks{e-mail: christian.mai@um.ifi.lmu.de}\\ %
     \scriptsize LMU Munich, Munich %
\and Thomas Kosch\thanks{e-mail: thomas.kosch@hu-berlin.de}\\ %
     \scriptsize HU Berlin, Berlin} %

\abstract{
When using Head-Mounted Displays (HMDs), users may not always notice or report visual discomfort by blurred vision through unadjusted lenses, motion sickness, and increased eye strain. Current measures for visual discomfort rely on users' self-reports those susceptible to subjective differences and lack of real-time insights. In this work, we investigate if Electroencephalography (EEG) can objectively measure visual discomfort by sensing Event-Related Potentials (ERPs). In a user study (N=20), we compare four different levels of Gaussian blur in a user study while measuring ERPs at occipito-parietal EEG electrodes. The findings reveal that specific ERP components (i.e., P1, N2, and P3) discriminated discomfort-related visual stimuli and indexed increased load on visual processing and fatigue. We conclude that time-locked brain activity can be used to evaluate visual discomfort and propose EEG-based automatic discomfort detection and prevention tools.
} 

\CCScatlist{
  \CCScatTwelve{Human-centered computing}{Visu\-al\-iza\-tion}{Visu\-al\-iza\-tion techniques}{Treemaps};
  \CCScatTwelve{Human-centered computing}{Visu\-al\-iza\-tion}{Visualization design and evaluation methods}{}
}

\begin{document}

\firstsection{Introduction}
\maketitle

Virtual Reality (VR) is becoming a mainstream technology that has been successfully employed in medicine \cite{vaughan2016review}, automotive applications \cite{lawson2016future}, and information visualization~\cite{donalek2014immersive}. However, the adoption of VR faces several challenges. \added{One frequently reported issue with asymmetric displays is blurred vision. For example, unadjusted lenses, motion sickness, and the vergence-accommodation conflict (VAC)~\cite{hoffman2008vergence}, a phenomenon responsible for visual discomfort and psychophysical strain, can lead to increased eye strain and asthenopia \cite{lambooij2009visual, bando2012visual,ukai2008visual}. Various studies showed that immersive content displayed in Head-Mounted Displays (HMDs) had been perceived as uncomfortable when the image was blurred \cite{kooi1997visual, morphew2004helmet, sharples2008virtual}, thus potentially hindering VR adoption.}



\added{In this study, we investigate if cortical activity indicates the perception of blurred vision in VR, using Electroencephalography (EEG) as a brain sensing technique. With the increasing trend to include physiological sensing into immersive environments~\cite{10.1145/3604270, 10.3389/frvir.2021.694567,10.1145/3319499.3328230}, EEG offers a significant advantage in evaluating the usability of HMD systems~\cite{10.1145/3229093} by providing objective measures for users' demands of cognitive~\cite{kosch2023cognitveworkloadhci} and visual workload~\cite{lorenc2021distraction}. This approach avoids potential biases often associated with subjective evaluations and a lack of real-time capabilities, such as revealing usability issues that users might not consciously notice during user testing and interaction with HMDs~\cite{10.1145/2858036.2858525}. Four levels of Gaussian blur were applied to VR scenes and used to explore general discomfort from visual distortions. This approach allows us to investigate the broader implications of visual discomfort in VR when users view blurred content using EEG as an objective metric.}

A growing body of research demonstrates EEG-based methods' efficacy in delivering more sensitive assessments than traditional behavioral and subjective evaluations. These methods have proven particularly effective in evaluating audio quality \cite{antons2012analyzing}, visual discomfort~\cite{scholler2012toward, lindemann2011evaluation}, and mental workload~\cite{mustafa2012eeg, 10.1145/3313831.3376766}. Initially focused on medical applications, techniques from brain-computer interfaces research \cite{lecuyer2008brain, putze2020brain} are now being increasingly applied in areas such as usability evaluation and the development of adaptive HMD applications \cite{wenzel2015eeg}. Previous work investigated if and how visual discomfort can be assessed using cortical activity measurements. This includes Event-Related Potentials (ERPs), a measured brain response from specific sensory, cognitive, or motor events.




The overall effect on visual fatigue was studied with EEG comparing 3D flat and panoramic screens versus HMD \cite{choy2021quality}, 2D versus 3D displays in HMDs \cite{kim2011eeg} and a short versus a long viewing duration \cite{li2008measurement}. Moreover, through measuring ERPs, Cho et al. \cite{cho2012feasibility} showed increased resource allocation for 3D asymmetric image processing, allowing for reliable visual discomfort classification. In the context of an adaptive system, Frey et al. \cite{frey2014assessing} classify EEG signals and evaluate visual comfort on a single-trial basis, allowing for accurate classification between uncomfortable and comforting conditions. These findings show how EEG is a feasible input for developing adaptive systems that adjust the stereoscopic experience to individual viewer preferences. This approach detects visual discomfort in real-time without relying on baseline recordings pre- and post-exposure EEG evaluations. For a more accurate, real-time assessment, it is thus crucial to expose subjects to shorter stimuli and immediately record electrophysiological responses, a gap our research aims to fill.

In this paper, we evaluate the efficiency of ERPs as a correlational measure for assessing different magnitude levels of \added{visual discomfort}. Inspired by previous work~\cite{mai2018evaluation}, we manipulate blur levels on either both HMD lenses (i.e., symmetrical blur) or one HMD lens at the same time (i.e., asymmetrical blur). In a user study (N=20), participants viewed four blur settings while measuring EEG for estimating ERPs. \added{We use Gaussian blur to simulate visual discomfort, varying the symmetrical and asymmetrical intensity across different trials to assess its impact on ERP responses.} Our result significantly affects the P1, N2, and P3 ERPs when viewing VR content distorted by \added{Guassian blur}. Our results are supported by the Simulator Sickness Questionnaire (SSQ)~\cite{balk2013simulator} and visual discomfort questionnaire~\cite{sheedy2003all}. We discuss the implications of our results and present how EEG can be integrated into common HMDs. We envision that the integration of EEG allows real-time sensing of visual discomfort to notify experimenters about needed adjustments in their experimental setups and use ERPs to inform designers about visual discomfort through blur in their applications.


\section{Related Work}
The following section introduces the main factors for visual discomfort in HMDs, explicitly focusing on visual discomfort.

\subsection{Visual Discomfort in HMDs}

Visual discomfort in the context of HMDs is a multifaceted issue, influenced by a variety of factors that strain the visual system, such as uncorrected visual errors and strenuous visual stimuli \cite{hirzle2020survey, hirzle2022understanding}. Among these, flicker caused by rapid changes in brightness or color is a significant concern. Particularly below the critical flicker fusion threshold, flickering becomes markedly perceptible, leading to eye strain and visual impairment and, in severe cases, to seizures \cite{palmer1999vision, wilkins2010led}. While advancements in display technology have mitigated mainly this issue, poorly designed color changes \cite{ishiguro2004follow}, and rendering artifacts, such as Z-fighting \cite{setthawong2015potential}, still pose problems. The susceptibility to flicker varies among individuals, influenced by factors like brightness, contrast, and physiological differences, with a notable increase in sensitivity within the peripheral vision \cite{wang2012luminance, wilkins2010led}. Moreover, image update latency is another critical factor contributing to visual discomfort. The discrepancy between actual head movement and the corresponding image update in HMDs can create a sensory mismatch, leading to discomfort and VR sickness, significantly if the latency exceeds 20\,ms \cite{long1994conformal, laviola2000discussion}.

The role of binocular asymmetries—encompassing both photometric differences such as contrast, brightness, color, and geometric variations such as shape, scaling, or rotation cannot be overstated in their contribution to visual discomfort \cite{gavrilescu2015visual, bando2012visual}. Geometric disparities, particularly vertical disparity, are known to cause discomfort within short viewing periods \cite{yamanoue1998tolerance, devernay2010stereoscopic}. Similarly, excessive horizontal disparity, crucial for creating asymmetric--symmetric imagery, becomes a source of discomfort when it is overly pronounced, challenging the visual system's ability to merge the two images into a cohesive visual experience. Research suggests maintaining specific thresholds for disparity to ensure comfortable viewing \cite{lambooij2009visual, shibata2011zone}.

Additionally, binocular rivalry, which occurs when significantly different images are presented to each eye, leads to a fluctuation in visual awareness that can cause eye strain and impair vision functionality. This effect is particularly pronounced with high-contrast images or when asymmetric techniques are improperly applied \cite{blohm1997stereoscopic}. Lastly, chromatic aberration, characterized by color fringes and blurred pictures at the boundaries of high-contrast areas, further contributes to the discomfort experienced with HMDs. This issue is typically addressed through specialized lenses or software corrections \cite{morphew2004helmet, frey2014assessing}.

\subsection{Using EEG for Sensing Visual Discomfort}

Visual discomfort in stereo viewing results from inappropriate binocular disparity. Depth features of asymmetric images have thus been widely investigated. Here, EEG is a tool for measuring visual discomfort due to its high temporal resolution, non-invasiveness, and ability to capture cortical brain activity in response to visual stimuli directly \cite{zafar2015decoding}. 
A specific EEG-based method of particular interest is the Visual Evoked Potential (VEP), which evaluates the functional integrity of the visual pathways extending from the retina, through the optic nerves, to the visual cortex. VEPs are elicited by presenting visual stimuli and recording the brain's electrical response from electrodes placed over the occipital lobe.
Recent studies utilizing either VEPs or ERPs investigated the neural correlates of visual discomfort, focusing on ERP components P1, N2, and P3.  Negishi et al.~\cite{negishi2012vep} conducted a study on visual discomfort in asymmetric viewing, examining the latency of the P1 component evoked by checkerboard pattern reversal stimulation, both before and after visual tasks. Their findings revealed that the latencies of P1 were delayed following the tasks, irrespective of whether they were presented in 3D or real space. This observation suggests that P1 may indicate visual fatigue induced by vergence eye movements, yet it does not exclusively pertain to 3D visual stimuli. Long et al. \cite{long2022assessment} evaluated 3D visual discomfort via functional connectivity analysis, suggesting that alterations in P1 could reflect the brain's initial response to conflicting visual cues in asymmetric images. The N2 and P3 components were investigated by Wu et al. \cite{wu2020inhibition} found that larger N2 and smaller P3 amplitudes are indicative of visually induced motion sickness (VIMS) and lead to hypothesizing a similar pattern in response to blur or stereoscopic disparity-induced discomfort.  

However, the effects of blur and to which degree the disparity would evoke visual discomfort have not yet been well studied based on VEP. Hence, in this work, we investigate ERPs, specifically the components P1, N2, and P3, to understand the impact of visual discomfort on ERP correlates of visual processing, conflict monitoring, and resource allocation.

\section{User Study}
In this study, we evaluate how symmetrical and asymmetrical Gaussian blur affects visual discomfort and ERPs of visual processing. The goal of our user study is to answer the following research question:


\begin{itemize}
    \item[\textbf{RQ:}] \added{Do ERP responses discriminate between different symmetric and asymmetric Gaussian blur intensities?}
\end{itemize}

We utilized a within-subjects experimental design. The independent variables were \textsc{Blur} (three levels: \textit{Neutral}, \textit{Low}, and \textit{High}) and \textsc{Symmetry} (two levels: \textit{Asymmetrical} and \textit{Symmetrical}). Therefore, independent variables were manipulated using a $3\,\times\,2$ experimental design.

\subsection{Task}

\begin{figure*}[h]
    \centering
    \includegraphics[width=\textwidth]{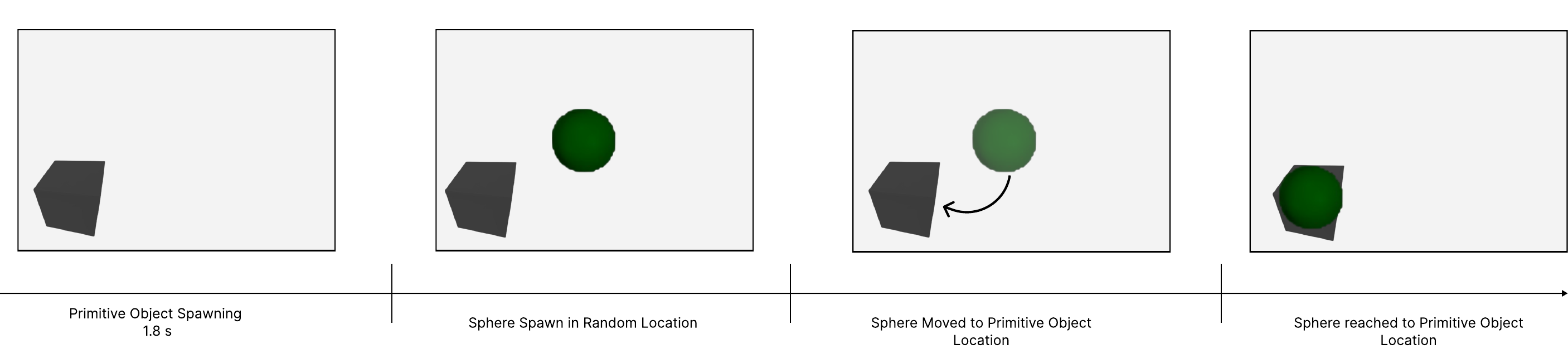}
    \caption{\emph{Experimental Task}. The task asked participants to interact with geometric stimuli. Initially, a primitive object appeared for 1.8 seconds and then disappeared. Subsequently, a sphere spawned at the center, and participants had to move it to where the object was last seen. This sequence was repeated with each new object appearing without any blur or with varying blur levels and symmetry (see \autoref{sec:ind_var}).
    }
    \label{fig:Task}
\end{figure*}

The task was a game where the participants viewed a geometric stimulus object based on previous work \cite{lochhead2022immersive, parsons2004sex} to maintain the attention of the participant (see \autoref{fig:Task}). The object disappeared after 1.8 seconds. Then, a sphere appeared in the center of the screen. The participant used the keyboard arrows to move the sphere to the position at which the last object appeared. Afterward, a new geometric object appeared, and the participants had to repeat the procedure. The Z-axis remained the same throughout the experiment. We decided that the participants would play a game since the lack of stimuli or proactive task was assessed negatively by users in previous studies~\cite{mai2017estimating}. The geometric stimulus object was either displayed without blur (i.e., baseline) or using one of the blur and symmetry levels mentioned before.



\subsubsection{Stimuli}

Mai et al. \cite{mai2018evaluation} evaluated through a survey of VR device and application developers which visual discomfort factors commonly occur in the everyday use of HMDs and how significantly they affect visual comfort. The survey results indicate blur, image update latency, and flickering \cite{KOOI200499, o2013visual}. Thus, in the design of the stimuli, we explored symmetrical and asymmetrical Gaussian blur. These visual conditions have been under-investigated regarding their electrophysiological impact, especially compared to reactions elicited by neutral stimuli. In HMDs, blurry images can be caused by low resolution or lens distortions when the eyes are not aligned with the lens center due to incorrect positioning of the HMD on the head or improper lens distance adjustment. If only one eye is misaligned, binocular asymmetries can occur, leading to blurred images.

The stimulus set consisted of representations of grey primitive 3D geometric shapes (Cube, Sphere, Capsule, Cylinder, RGB values: approx. 61,61,61 - dependent on viewing angle). The background was dark grey (RGB values: 41,41,41) to avoid reflections on the lenses. The blur effect is created using a Gaussian blur filter. Blurry representations in two gradations are displayed with binocular symmetry and binocular asymmetry, resulting in four conditions inducing visual discomfort. \added{It is essential to distinguish between the inherent blur applied to the objects in our experiments and the perceptual blur experienced as a result of VAC. While the former is a controlled visual condition used to elicit discomfort, the latter pertains to a viewer's subjective experience of visual clarity. Lambooij et al. \cite{lambooij2009visual} demonstrate that artificial blurring and VAC-induced perception issues elicit similar types of visual discomfort. Hence, we use Gaussian blur as a proxy to investigate the effects of visual discomfort on cognitive and perceptual processes in VR environments.}

In the asymmetric stimuli A1 and A2, the left or right image is randomly blurred. We included asymmetric blur to ensure a discomfort-inducing effect, using empirically determined blur levels that reliably cause discomfort in users. The rationale behind this choice is to create a sensory mismatch, where the visual system receives two incongruent stimuli. This break in sensory integration mimics real-world scenarios where users might experience visual strain due to misaligned lenses or other visual discrepancies in VR settings.  The choice of the Sigma ($s$) value as a measure of blur is based on empirically determined blur levels that cause discomfort in users. Here, an $s$ of $1.0$ pixel and $s$ of $1.5$ pixels showed suitable values for inducing discomfort on binocular displays~\cite{kooi2004visual}. In contrast, we decided to ensure a discomfort-inducing effect of the stimuli by choosing a $s$ of $2.0$ pixels and $4.0$ pixels, respectively. We include a neutral stimulus without any blur as a baseline. Overall, the study included five conditions.

\begin{figure}[h]
    \centering
    \includegraphics[width=\columnwidth]{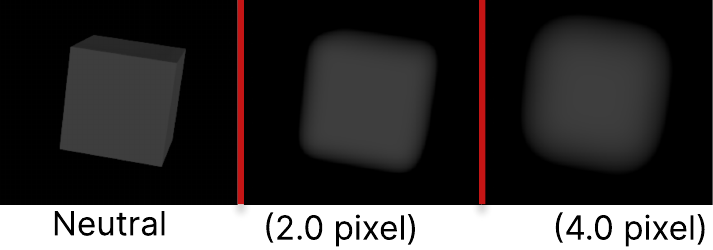}
    \caption{\emph{Experimental Stimuli for the \textsc{Blur Level}}.}
    \label{fig:stimuli}
\end{figure}

\begin{figure}[h]
    \centering
    \includegraphics[width=\columnwidth]{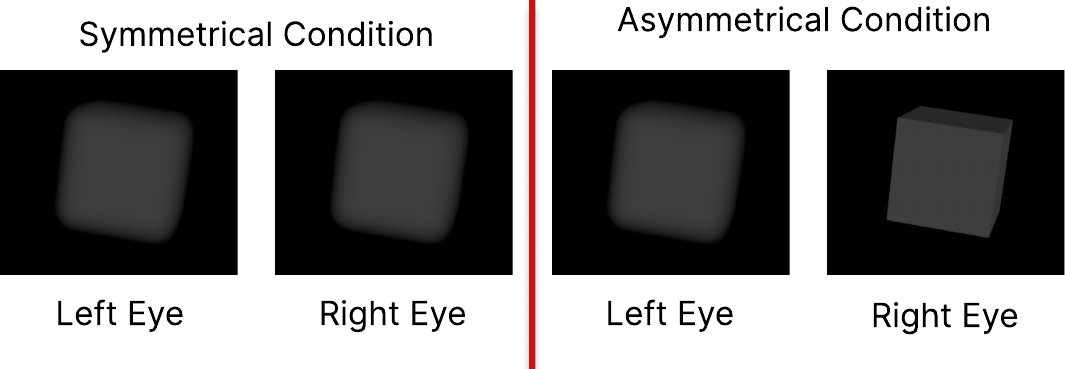}
    \caption{\emph{Experimental Stimuli for the \textsc{Symmetry Adjustment}}.}
    \label{fig:asymmetric_stimuli}
\end{figure}

\subsection{Independent Variables}
\label{sec:ind_var}
In our study, we use the \textsc{Blur Level} and the \textsc{Symmetry Adjustment} as independent variables. 

The \textsc{Blur Level} depicts the degree of artificially induced blur recommended by previous research to simulate visual discomfort~\cite{kooi2004visual}. The \textsc{Blur Level} includes the following two levels \textit{Slight Blurriness} ($s$ = $2.0$) and \textit{Severe Blurriness} ($s$ = $4.0$). The \textsc{Symmetry Adjustment} indicates whether the stimulus is shown with the same or different blur levels on each lens (see \autoref{fig:stimuli} and \autoref{fig:asymmetric_stimuli}). 

The \textsc{Symmetry Adjustment} includes the two levels \textit{Symmetric} and \textit{Asymmetric}. Finally, we provide a sharp stimulus without blur or symmetry adjustments as a baseline. Overall, the experimental designs include five different stimuli. We provide an overview of the experimental levels in~\autoref{tab:conditions}.

\begin{table}[]
\centering
\caption{Conditions used in the user study. We vary the \textsc{Blur Level} using a Gaussian blur filter and \textsc{Symmetry Adjustment} for the HMD lenses to simulate visual discomfort.}
\label{tab:conditions}
\begin{tabular}{@{}lll@{}}
\toprule
\textbf{Condition} & \textbf{Blur Level in s} & \textbf{Symmetry Adjustment} \\ 
\midrule
N                  & 0.0                     & None                         \\
S1                 & 2.0                      & Symmetric                 \\
S2                 & 4.0                      & Symmetric                \\ 
A1                 & 2.0                      & Asymmetric                \\
A2                 & 4.0                      & Asymmetric               \\
\bottomrule
\end{tabular}
\end{table}




\subsection{Dependent Variables}

We collected EEG data throughout the experiment. The beginning of each trial was marked in the EEG data using a COM-Port emulator\footnote{\url{https://eterlogic.com/Downloads.html}} for later extraction of each epoch. \added{Specifically, we investigate three ERP components, i.e., P1, N2, and P3, for evaluating early visual processing, conflict detection, and resource allocation, providing a detailed understanding of how visual discomfort influences cognitive functions in real time. \cite{gaspelin2018combined}. 
Unlike frequency-based analysis, ERPs allow us to dissect complex cognitive responses impacted by visual discomfort.} 
Furthermore, we measured the subjectively perceived discomfort by utilizing the visual discomfort \cite{sheedy2003all} and the Simulation Sickness Questionnaire (SSQ) \cite{balk2013simulator}.

\subsection{Procedure}

\begin{figure*}[h]
    \centering
    \includegraphics[width=\textwidth]{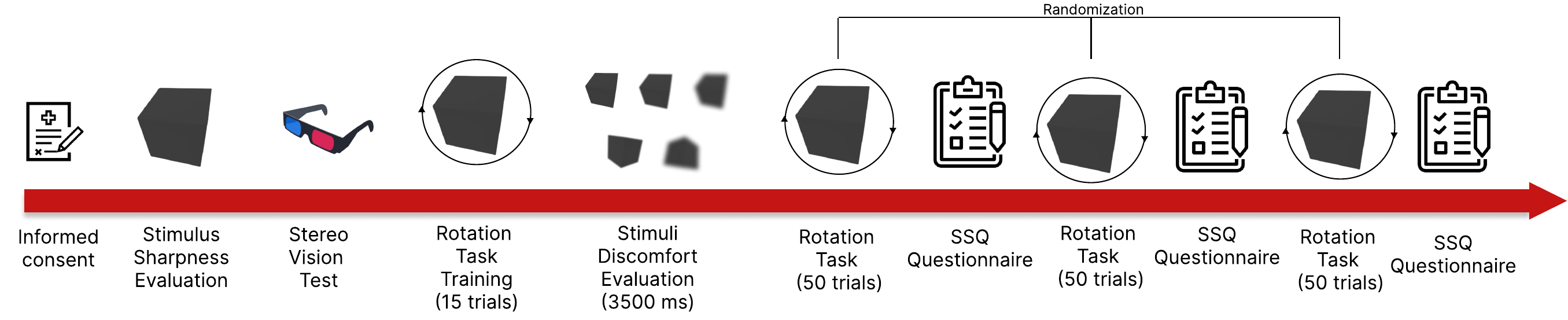}
    \caption{\emph{Experimental Procedure}. After obtaining informed consent, participants underwent various preliminary assessments, including stimulus sharpness evaluation, stereo vision test, and initial SSQ questionnaire. They then engaged in a series of rotation tasks divided into training (15 trials) and three main sessions (50 trials each), with trials randomized across conditions. Between sessions, participants evaluated stimuli, discomfort, and sharpness. The experiment concluded with a final SSQ questionnaire to reassess simulator sickness.}
    \label{fig:Trial_AR}
\end{figure*}

Upon the participants' arrival, the experimenter provided them with details about the study's process and obtained informed informed consent—afterward, the participants filled in their demographic data. We measured the interpupillary distance and darkened the room. The participants were seated on a chair. The participants were then put on the adjusted HMD, which was adjusted to the participant's interpupillary distance, and the EEG headset. The participants were initially viewing a grey cube on a black background. The experimenter assessed if the EEG signal was noisy. If the EEG signal was noisy, the experimenter adjusted the EEG headset or added additional saline solution to improve the conductivity. During the experiment, the participants were instructed to minimize eye blinks and head movements to reduce noisy artifacts through muscular movements. Afterward, the participants were asked if the stimulus appeared sharp or if further HMD adjustments were necessary, such as adjusting the HMD lenses according to the measured interpupillary distance. We tested the stereo vision of the participants and measured the interpupillary distance. Then, the participants orally answered the SSQ~\cite{balk2013simulator}. \autoref{fig:Trial_AR} illustrates the study procedure.

The participants played 15 test trials of the rotation task game using sharp cubes only to get acquainted with the task. The experiment continued when the participants confirmed they were comfortable with the task. The participant then sequentially viewed five cubes with each blur and symmetry level for 3.5 seconds at the center of the HMD on a gray background in random order. 
\added{To mitigate the potential cumulative effects of visual strain over continuous viewing periods, we integrated breaks after every 50 trials to allow participants' eyes to rest. This approach is designed to minimize the bleed-over effects of blur and asymmetries or any residual discomfort from the experimental conditions. These breaks ensure that any subsequent measurements are not unduly influenced by prior exposure, thereby maintaining the integrity of the data collected. This protocol adjustment also helps to differentiate the immediate effects of our induced visual conditions from longer-term adaptation or fatigue effects that might otherwise skew the results.}

We recorded the EEG data as a baseline for each viewed cube and asked the participant to answer the visual discomfort questionnaire~\cite{sheedy2003all} verbally. A sharp cube was displayed while answering the questionnaire to avoid participant discomfort by remaining in an empty scene~\cite{mai2017estimating}. The participants repeated the procedure until they had seen all five cubes. 

A total of three sessions were conducted per participant. A session displayed each condition in a random order ten times, resulting in 50 trials per run. Each stimulus was displayed for 1800 milliseconds. Afterward, participants could conduct a break without putting the EEG and the HMD off. A sharp cube was shown during the break. The next session began when participants confirmed they would continue. Overall, the participants conducted 150 trials during the whole study. Finally, the participant answered the SSQ again and removed the EEG and the HMD, concluding the experiment. \added{Overall, each participant engaged in 150 trials, distributed across five conditions (30 trials per condition), totaling 3000 stimuli presentations (600 per condition). This data collection aligns with the recommended trial amount \cite{boudewyn2018many}, allowing for robust statistical power.}  The total duration of the experiment, including briefing and setup, was 60 minutes.



\subsection{Apparatus}
We implemented the task in Unity (Version 2017.4.1f1) and presented the stimuli using an Oculus Rift headset (1080 × 1200 pixels, field of view: 110\textdegree). We acquired EEG signals using EMOTIV EPOC+\footnote{\url{www.emotiv.com/products/epoc}}. EEG data were streamed within the Unity VR environment within the Lab Streaming Layer (LSL) framework\footnote{\url{www.github.com/labstreaminglayer}} to the acquisition PC (Windows 10, Intel Core i7-11700K, 3.60 GHz, 16GB RAM). 

\subsection{EEG Preprocessing and ERP Analysis}
EEG data were collected using an Emotiv EPOC+ EEG cap equipped with 16 water-based electrodes (AF3, F7, F3, FC5, T7, P7, O1, O2, P8, T8, FC6, F4, F8, AF4). M1 and M2 were used as reference electrodes according to a 10-20 layout. \added{Emotiv EPOC+ has been validated for its accuracy in measuring auditory ERPs  \cite{badcock2013validation} and has demonstrated sufficient sensitivity in visual processing ERPs, including N2 and P3\cite{de2015measuring, balart2019step}. Specifically, the electrodes positioned at O1 and O2 have been previously validated for effectively discerning visual stimuli differences \cite{clayson2021data}.} We ensured that the electrode impedance was maintained below 20 k$\Omega$. The LSL framework allowed for time synchronization with the Unity environment. 

Data preprocessing and analysis were conducted using the MNE-Python Toolbox~\cite{gramfort2013meg}. The signal was notch-filtered at 50 Hz and band-pass filtered between 1-15 Hz to remove noise. \added{We selected such filter settings based on ERP guidelines \cite{clayson2021data, widmann2015digital,govaart2022eeg} to optimize noise removal while preserving the integrity of important ERP component parameters, such as peak amplitudes.}

We then re-referenced the electrodes of our EEG measurements to the average of all electrodes. Then, we conducted an Independent Component Analysis (ICA) using the extended Infomax algorithm for artifact detection and correction~\cite{lee1999independent}. The ICLabel MNE plugin was used for automated labeling and rejection of ICA components~\cite{li2022mne}. Epochs contaminated with blinks, eye movements, muscle artifacts, or single-channel artifacts were excluded from the analysis. 
For ERP analyses, we segmented continuous signals between 200 ms before and 1000 ms after the stimulus presentation, removing a 200 ms baseline before stimulus onset. Electrodes of interest were those over early visual areas, P7, P8, O7, and O8. As these showed a similar pattern of results, data were averaged over the four channels \cite{o2023relationship}. On average, we removed $.93$ ($SD=.5$) independent components within each participant. We analyzed the three components: P1, N2, and P3. The P1 and P3 were quantified as positive average peak amplitudes in the 100 -- 300 ms and 300 -- 600 ms windows \cite{stormer2013normal}, respectively, while for the N2 component, we computed the average negative peak in 100–300 ms time window \cite{righi2009anxiety}. These windows were centered upon the peak latency of each component in the grand average waveforms \cite{roche2005individual}.

\subsection{Participants}
We recruited 20 participants ($M=25.4$, $SD=4.8$; 8 self-identified as female, 12 self-identified as male). Among them, 55\% (11 of 20) were first-time VR users. \added{\added{Our recruited sample size for our study follows recent investigations on the relationship between participants number and EEG data reliability for relatively long tasks~\cite{vozzi2021sample} and in line with previous work in HCI \cite{long2024multimodal}, visual tasks ~\cite{chiossi2024impact} and physiological computing domains \cite{chiossi2023designing}.}} Additionally, 30\% (6 of 20) were playing video games, engaging in at least ten gaming sessions monthly. None had a history of neurological, psychological, or psychiatric symptoms, and all participants met the requirement for perceiving asymmetric images with a disparity of up to 100 arcsec. 

\subsection{Results}
\begin{figure}[t]
    \centering\includegraphics[width=\columnwidth]{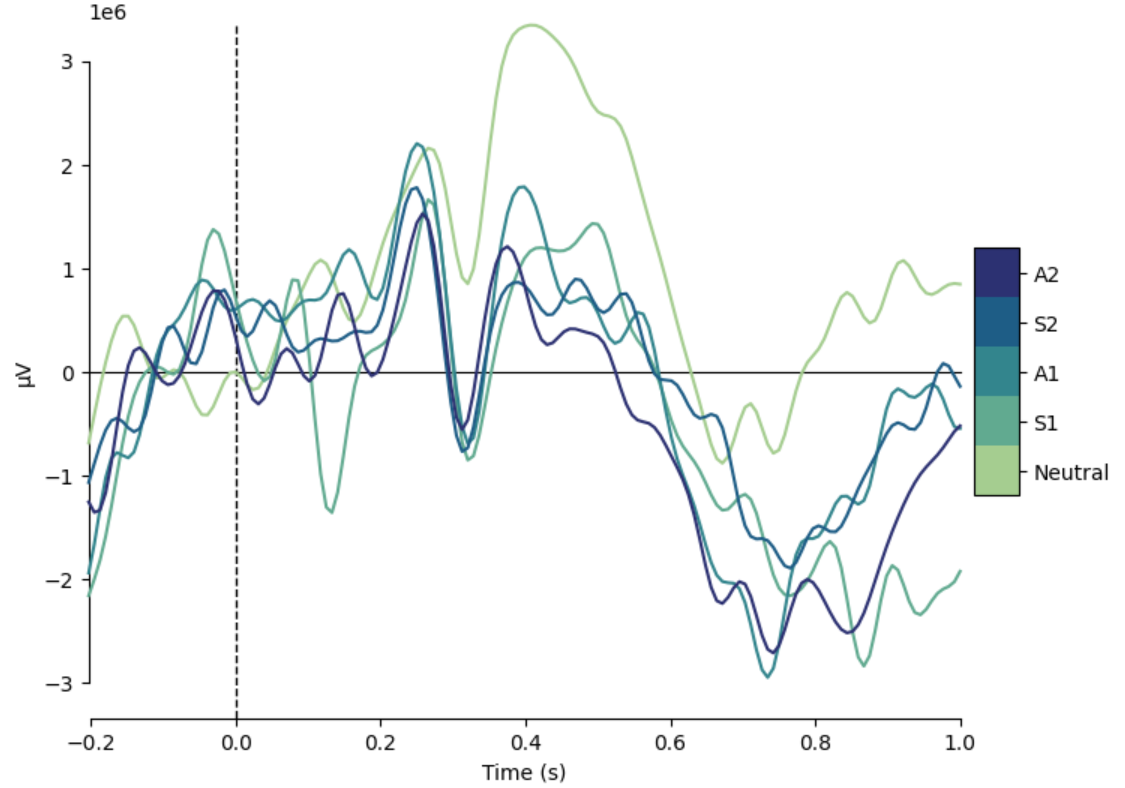}
    \caption{\emph{ERP for all conditions.} We display ERP waveforms from - .2 to 1.0 seconds relative to stimulus onset across different visual discomfort conditions, highlighting variations in P1, N2, and P3 components under various levels of blur and symmetry. The line plot illustrates differential impacts on ERP amplitudes: P1 (100-300 ms) shows minimal change across conditions, suggesting resilience in early visual processing; N2 (100-300 ms) amplitude decreases with increased blur, indicating enhanced conflict monitoring; P3 (300 - 600 ms) demonstrates varied responses, with subtle reductions suggesting a marginal impact on cognitive resource allocation under altered visual conditions. }
    \label{fig:ERP_all}
\end{figure}

\begin{figure}[t]
    \centering
    \includegraphics[width=\columnwidth]{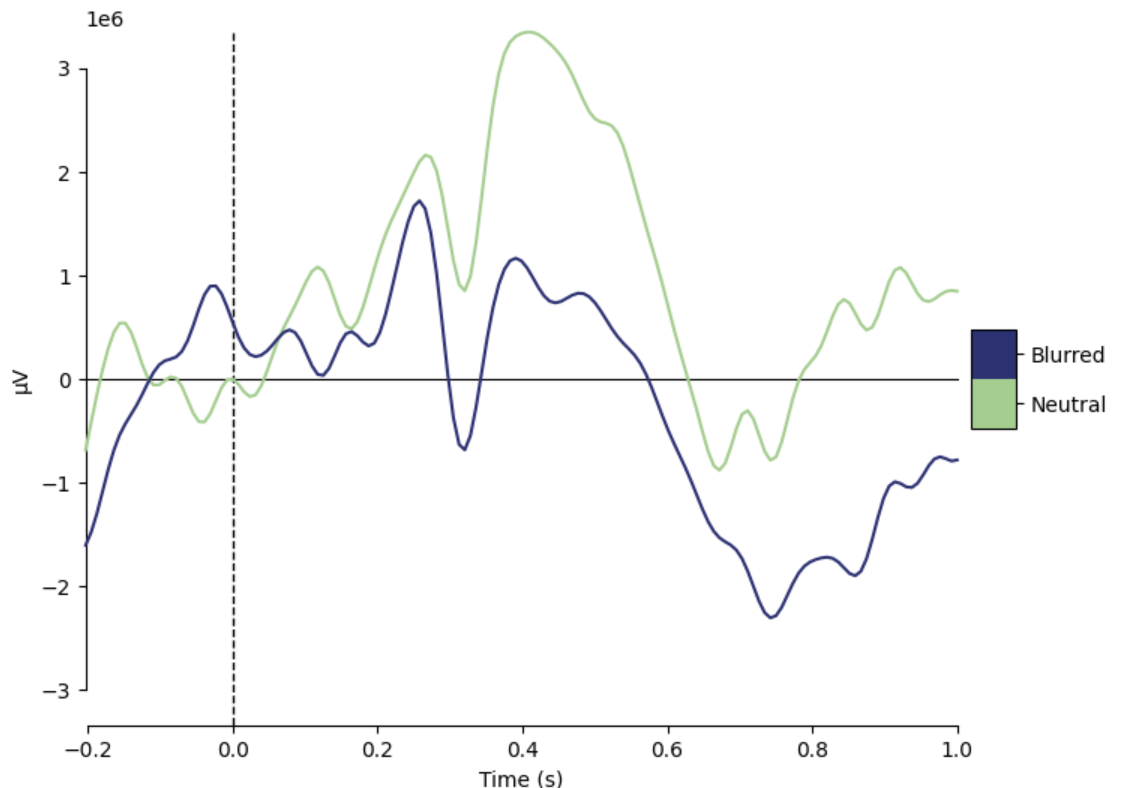}
    \caption{\emph{ERP for Blurred vs Neutral conditions}. We depict ERP for Neutral and Blurred visual conditions, illustrating the effects of visual discomfort on the P1, N2, and P3 components. The results highlight significant differences: the P1 component shows increased amplitude in the Neutral condition, suggesting enhanced early sensory processing; the N2 amplitude is reduced under Blurred conditions, indicating greater conflict monitoring; and the P3 component exhibits higher amplitudes in Neutral settings, reflecting more efficient cognitive resource allocation. These findings demonstrate how blur-related visual discomfort impacts cognitive processing in VR environments.}
    \label{fig:ERP_average}
\end{figure}

\begin{figure*}[h]
    \centering
    \includegraphics[width=\linewidth
    ]{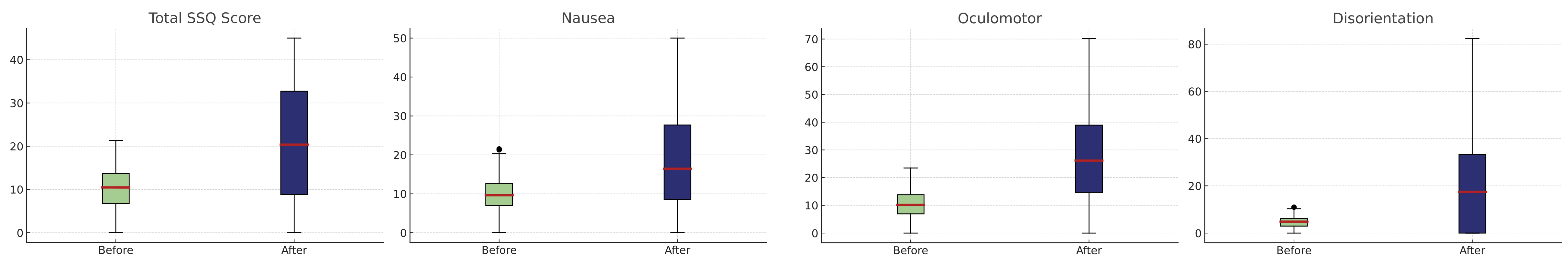}
    \caption{\emph{SSQ Results}. Participants were asked to report their subjective simulator sickness before and after the end of the experiment. They reported increased simulator sickness at a subscale level and aggregated SSQ score. Results indicate a noticeable increase in all measures post-exposure, with the Total SSQ Score and specific symptoms like Oculomotor and Disorientation showing significant escalation, reflecting heightened levels of simulator sickness induced by the experimental manipulations.}
    \label{fig:SSQ_results}
\end{figure*}

\begin{figure*}[h]
    \centering
    \includegraphics[width=.8\linewidth]{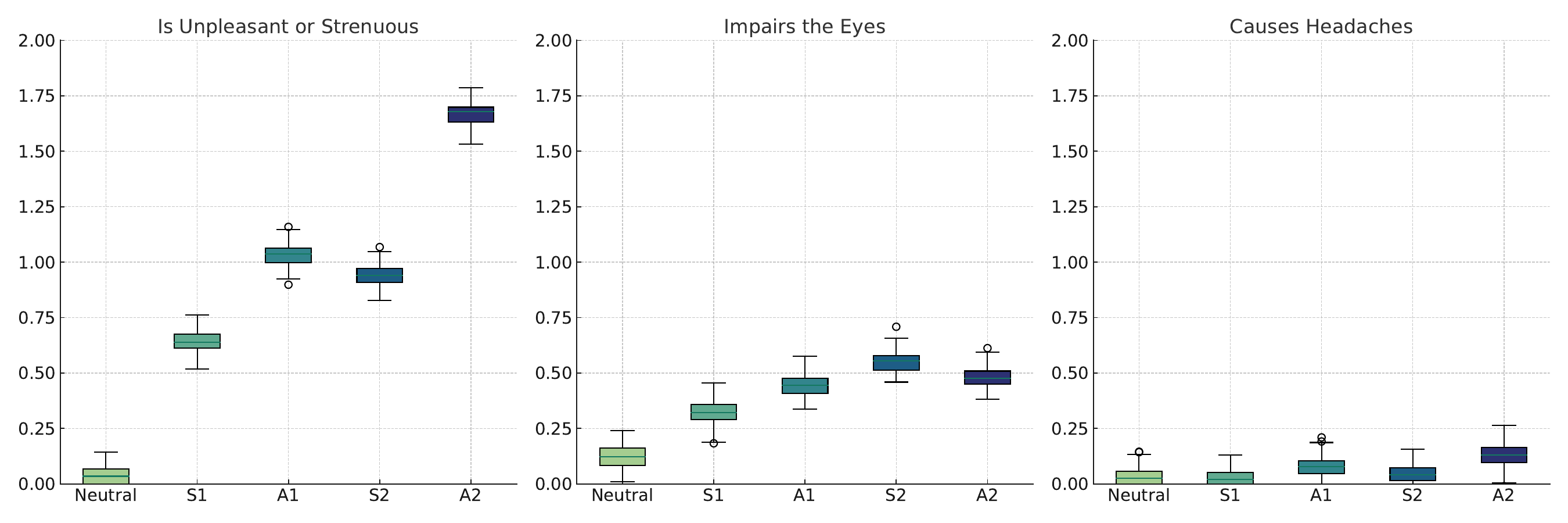}
    \caption{\emph{Visual Discomfort Results.} We depict participant ratings of discomfort related to unpleasantness, eye strain, and headaches across conditions (Neutral, S1, A1, S2, A2). Results indicated significant increases in discomfort from the Neutral condition, with A2 consistently rated as the most discomforting across all categories. Notable differences among other conditions suggest varying impacts on visual discomfort, with S2 prominently affecting eye strain and A2 heightening headache reports.
    }
    \label{fig:discomfort_results}
\end{figure*}

In this section, we first present results on ERP analysis on peak amplitude for P1, N2, and P3 across conditions and by aggregating the blurred stimuli vs the baseline condition. We employ a Generalized Linear Mixed Model (GLMM) to investigate differences in the ERP and subjective scores (formula: \texttt{Measure $\sim$ Blur * Symmetry + (1|participant)}). We utilized a restricted maximum likelihood estimation (REML) with the nloptwrap optimizer. For subjective scores on SSQ and perceived discomfort, we use either t-test pairwise comparisons or Repeated Measures ANOVA.~\autoref{fig:ERP_all} and~\autoref{fig:ERP_average} illustrate the results of the ERPs.~\autoref{fig:SSQ_results} and~\autoref{fig:discomfort_results} visualize the results of the questionnaires.

\subsubsection{ERPs}

\paragraph{P1}

The model for predicting P1 amplitude with Blur and Symmetry yielded a conditional $R^2 = .40$, indicating that 40\% of the variance in P1 amplitude could be attributed to the model, with a marginal $R^2 = 9.96e-03$, suggesting the fixed effects of Blur intensity and Symmetry alone explained a minor portion of the variability in P1 amplitude. The model’s intercept for the baseline condition of Neutral Blur intensity and Symmetrical Symmetry was significantly high ($B = 5.66$, $95\%$ CI $[3.82, 7.50]$, $t(74) = 6.14$, $p < .001$). Despite the varying levels of Blur intensity and Symmetry, the effects on P1 amplitude were found to be non-significant: Slight Blur intensity ($\beta = -1.01$, $95\%$ CI $[-3.04, 1.02]$, $t(74) = -1.00$, $p = .323$), Strong Blur intensity ($\beta = -.37$, $95\%$ CI $[-2.40, 1.65]$, $t(74) = -.37$, $p = .714$), and Asymmetrical Symmetry ($\beta = .36$, $95\%$ CI $[-1.67, 2.39]$, $t(74) = .35$, $p = .725$).  These results indicate that variations in blur intensity and Symmetry do not significantly influence the amplitude of the P1 component, suggesting that early visual processing, as reflected by the P1 component, may be relatively resilient to visual discomfort manipulations.

\paragraph{N2}
When predicting N2 peak amplitude, we found significant adverse effects for both levels of Blur intensity. Low Blur intensity exhibited a significant decrease in N2 amplitude ($\beta = -2.12$, $95\%$ CI $[-3.96, -.29]$, $t(74) = -2.30$, $p = .024$), and High Blur intensity similarly led to a reduction in amplitude ($\beta = -1.98$, $95\%$ CI $[-3.82, -.14]$, $t(74) = -2.15$, $p = .035$). The effect of Symmetry was not significant and positively oriented ($\beta = .70$, $95\%$ CI $[-1.13, 2.54]$, $t(74) = .76$, $p = .449$). These results indicate that increases in Blur intensity, regardless of whether they are Slight or Strong, significantly dampen the N2 amplitude, suggesting an increase in cognitive effort or conflict monitoring as visual clarity decreases. Conversely, the alteration in Symmetry does not substantially impact N2 amplitude, implying that the mental processes reflected by the N2 component are more sensitive to changes in visual clarity than to alterations in pattern symmetry.

\paragraph{P3}
The model for P3 amplitude prediction demonstrated a significant conditional $R^2 = .58$, indicating that 58\% of the variability in P3 amplitude was accounted for by the model, with the fixed effects alone explaining a marginal 2\% of the variance (marginal $R^2 = .02$). In this model, the impact of Blur intensity and Symmetry on P3 amplitude did not show significant results. Slight Blur intensity produced a non-significant negative effect on P3 amplitude ($\beta = -.94$, $95\%$ CI $[-2.54, .66]$, $t(74) = -1.17$, $p = .246$), and Strong Blur intensity similarly showed a non-significant negative influence ($\beta = -1.50$, $95\%$ CI $[-3.10, .11]$, $t(74) = -1.86$, $p = .067$). The alteration from Symmetrical to Asymmetrical Symmetry resulted in a non-significant positive effect ($\beta = .27$, $95\%$ CI $[-1.33, 1.87]$, $t(74) = .33$, $p = .739$), suggesting that changes in visual symmetry had a minimal impact on P3 amplitude. These findings imply that the P3 component is resistant to variations in visual clarity and symmetry alone, with a general trend indicating that increased Blur intensity marginally decreases P3 amplitude.

\subsubsection{Neutral vs. Blurred ERPs}

\paragraph{P1}
In examining the impact of visual discomfort on the averaged P1 component amplitude, we found a significant difference between the Blurred and Neutral conditions. The model revealed a conditional \(R^2 = .66\), indicating a strong relationship between condition and P1 amplitude, with the fixed effect of Condition alone explaining 10\% of the variability (marginal \(R^2 = .10\)). Specifically, the intercept for the Blurred condition was significant (\(B = 2.83\), \(95\% CI [1.67, 3.99]\), \(t(36) = 4.95\), \(p < .001\)), and transitioning to the Neutral condition significantly increased the P1 amplitude (\(\beta = 1.70\), \(95\% CI [.70, 2.71]\), \(t(36) = 3.43\), \(p = .002\)). This finding indicates that visual discomfort associated with blur significantly modulates early sensory processing, as reflected in the P1 component. 

\paragraph{N2}
In assessing the impact of visual discomfort on the average N2 component amplitude, significant differences emerged between the Blurred and Neutral conditions. The model's conditional \(R^2 = .47\) and marginal \(R^2 = .29\) indicate a strong relationship between experimental conditions and N2 amplitude. The intercept for the Blurred condition, though not significant (\(B = -1.10\), \(95\% CI [-2.41, .21]\), \(t(36) = -1.70\), \(p = .098\)), serves as a basis for comparing the effect of the Neutral condition. Notably, N2 responses to the Neutral condition significantly decreased the average amplitude (\(\beta = 3.66\), \(95\% CI [2.06, 5.27]\), \(t(36) = 4.63\), \(p < .001\); \(Std. beta = 1.08\), \(95\% CI [.60, 1.55]\)) as compared to the blurred condition. This decrease in N2 amplitude for the Blurred condition shows that increased cognitive resources are needed for conflict monitoring in visual information when exposed to blurred images.

\paragraph{P3}
The model for predicting P3 amplitude demonstrated substantial explanatory power (conditional \(R^2 = .73\), marginal \(R^2 = .27\)), with the Neutral condition significantly increasing P3 amplitude (\(\beta = 3.85\), \(95\% CI [2.60, 5.10]\), \(t(36) = 6.24\), \(p < .001\)) compared to the Blurred condition (\(B = 4.51\), \(95\% CI [3.06, 5.97]\), \(t(36) = 6.30\), \(p < .001\)).
This decrease suggests that visual discomfort stemming from blurred stimuli directly impairs cognitive resource allocation, posing a mental strain on users.

\subsubsection{Simulator Sickness Questionnaire}
In assessing the effects of the symmetric and asymmetric blur manipulation on simulator sickness, scores on the SSQ scores were analyzed before and after the study. The pre-study mean score for Nausea was $12.88$ ($SD = 7.4$), which increased to a post-study mean of $18.60$ (SD = $11.1$); however, this increase was not statistically significant, p = $.155$. A substantial rise in symptoms was observed in the Oculomotor subscale, with the mean score escalating from $9.85$ (SD = $7.4$) pre-study to $26.91$ (SD = $37.0$) post-study, p $<.001$. Similarly, the Disorientation subscale showed a substantial increase from a pre-study mean of $9.05$ (SD = $.12$) to a post-study mean of $27.84$ (SD = $29.6$), p = $.002$. The Total SSQ score also significantly increased from a pre-study mean of $5.08$ (SD = $7.4$) to a post-study mean of $12.98$ (SD = $14.8$), p = $.002$.

\subsubsection{Visual Discomfort Questionnaire}

A Wilcoxon signed-rank test was utilized to compare the central tendencies of discomfort ratings among various stimuli. Results revealed significant contrasts. For the \textsc{Pleasantness},  neutral stimulus N was associated with the lowest discomfort (\(M = .06\), \(SD = .10\)). In contrast, the A2 stimulus was rated as the most discomforting (\(M = .76\), \(SD = .43\); \(p < .0001\)). Other stimuli demonstrated significant differences when compared to N, with S1 (\(M = .65\), \(SD = .24\); \(p < .0001\)), A1 (\(M = 1.04\), \(SD = .37\); \(p < .0001\)), and S2 (\(M = .93\), \(SD = .41\); \(p < .0001\)) all eliciting greater discomfort. Regarding \textsc{Eye impairment}, further comparisons using the Wilcoxon signed-rank test indicated that A2 was perceived as significantly more discomforting than S1 (\(p = .001\)) and A1 (\(p < .0005\)). Lastly, when comparing reports on \textsc{Causing Headaches}, significant differences were also observed between S1 and S2 (\(p = .0049\)), as well as between S2 and A2 (\(p = .0360\)), confirming significant differences in the levels of discomfort provoked by the discomforting stimuli.


\section{Discussion}

This study evaluated the effects of symmetry and blur intensity on specific ERP components, including P1-, N2-, and P3—correlates of perceptual processing, conflict monitoring, and resource allocation. Participants were exposed to a series of stimuli designed to elicit varying levels of visual discomfort and a neutral reference stimulus within a controlled within-participants experimental design. Here, we discuss our results on the impact of our dependent variables on ERP components and subjective scores. Then, we explore implications for the development of discomfort detection and the design of adaptive systems to optimize user interaction by mitigating visual discomfort.

\subsection{Impact of Blur Intensity and Symmetry on Visual Processing and Discomfort}
\label{sec:disc_P1_N2}

Results on P1, N2, and P3 components under conditions of varying \textsc{blur levels} and \textsc{symmetry adjustments} showed different results informative on visual perceptual and cognitive processing mechanisms in response to visual discomfort. The P1 component, reflecting early visual processing, was resilient to blur and symmetry pattern changes. This suggests that the early stages of visual processing appear to be relatively unaffected by variations in blur and symmetry. This implies that the initial sensory processing of visual stimuli is robust against such manipulations, focusing more on essential feature detection that precedes detailed analysis or recognition. Here, the brain's early visual processing system is highly efficient in handling visual information, even when the stimuli are blurred or asymmetric. This efficiency could result from the brain's ability to fill in gaps or ignore certain imperfections in visual input to maintain rapid processing speeds \cite{rousselet2011modelling}. Such findings align with Rousselet et al. \cite{rousselet2011modelling}, who discuss the modulation of bottom-up visual processing by top-down task constraints, indicating that early visual processing can adapt to various conditions without significant detriment to the processing efficiency.

In contrast, the N2 component, associated with cognitive control and conflict monitoring, showed a significant decrease in amplitude when exposed to variations in \textsc{blur levels}, for both low and high blurring resulting. This significant response indicates that visual discomfort, mainly from blurred stimuli, necessitates increased cognitive effort for effective processing. N2 response to blurred stimuli reflects an intermediate stage of cognitive processing where the brain recognizes and attempts to reconcile the discrepancy or conflict induced by visual discomfort \cite{wu2020inhibition}. The more negative amplitude observed in response to blurred stimuli indicates that the brain allocates additional cognitive resources to process these stimuli effectively, reflecting an effortful engagement to overcome the visual discomfort and maintain task performance. This is consistent with the literature suggesting that the N2 component reflects processes related to cognitive control and the mobilization of cognitive resources in response to challenging or unexpected stimuli \cite{folstein2008influence}. Furthermore, the lack of significant change in N2 amplitude in response to symmetry alterations highlights the component's specificity in dealing with challenges to visual clarity rather than pattern organization. This specificity aligns with the understanding that cognitive control mechanisms, as indexed by N2, are remarkably tuned to address discrepancies in expected versus actual sensory input, prominently challenged by blur \cite{gaspelin2018role}.

\subsection{Impact of Blur Intensity on Cognitive Resource Allocation}

The P3 component, reflective of cognitive engagement and resource allocation, showed a trend of decreased amplitude with heightened blur intensity, although this did not reach statistical significance. This pattern suggests a limit to the cognitive system's adaptability or the activation of compensatory mechanisms to counteract the effects of reduced visual clarity \cite{polich2007updating}. Despite the strain posed by the visual discomfort conditions, the slight modulation in P3 amplitude indicates an effort to maintain higher-order cognitive processes, highlighting a trade-off between adaptability and the mental demands of processing discomforting visual stimuli.

\subsection{Impact of Aggregated Blurred Conditions on Visual Processing}

As an integrative analysis, we aggregated  \textsc{Blur Level} and \textsc{Symmetry Adjustment} conditions to compare against a baseline Neutral condition offers to assess how deviations from clear, standard visuals—through blur or symmetry changes impact the ERP correlates of process related to visual strain. This combined approach is critical because, in everyday VR environments, individuals often encounter visual stimuli that vary in clarity and symmetry.

The findings from comparing Neutral and aggregated Blurred conditions reveal significant results.  The P1 component's decreased amplitude in Blurred conditions indicates an increased visual processing load.  This reduction in amplitude suggests that the initial stages of visual processing are taxed by the effort to resolve the ambiguity and reduced clarity inherent in blurred visuals. 
N2 showed decreased and more negative amplitudes for the aggregated conditions, aligned with individual analysis results (see \autoref{sec:disc_P1_N2}. Here again, increased N2 amplitude can be interpreted as indicative of heightened cognitive effort and conflict monitoring.  First, the attempt to decipher blurred stimuli, especially when symmetry is expected, challenges the visual system, leading to increased neural activity indicative of this processing demand \cite{folstein2008influence}. Second, a more negative N2 amplitude indicates enhanced conflict monitoring, where the brain actively detects and responds to discrepancies between expected and actual visual inputs \cite{yeung2004neural}. This enhanced monitoring reflects the cognitive system's attempt to reconcile the conflict induced by the blurred yet symmetric stimuli, a cognitively demanding process.
 
Lastly, we found a decrease in P3 amplitude observed in the Blurred condition, which suggests that visual discomfort resulting from blurred stimuli directly impacts cognitive resource allocation and processing efficiency. The P3 component, widely regarded as a marker of attentional resource allocation \cite{polich2007updating} and the processing of stimulus significance, informs the impact of blurred and asymmetrical stimuli on users' capacity to engage with and respond to environmental stimuli. A decrease in P3 amplitude in response to blurred stimuli indicates a reduction in the cognitive system's ability to allocate attentional resources effectively. This reduction can be attributed to the increased cognitive load imposed by the effort to interpret blurred images, which can detract from the cognitive resources available for processing other aspects of the stimuli, such as their significance or relevance to the task. Consequently, this diminished resource allocation can lead to decreased efficiency in cognitive processing. This interpretation aligns with existing literature that emphasizes the role of visual clarity in cognitive efficiency and task performance. For instance, studies have shown that transparent and easily interpretable visual stimuli facilitate better cognitive engagement and faster processing speeds, as reflected by higher P3 amplitudes \cite{polich2007updating}. Conversely, when visual stimuli are blurred, necessitating greater effort for interpretation, cognitive resources are stretched thin, leading to decreased P3 amplitude and, by extension, reduced cognitive efficiency \cite{kok2001utility}.

Such conditions, overall, show how \textsc{Blur Level} and \textsc{Symmetry Adjustment} place demand on users to perform essential feature detection and sensory encoding, indicating that the brain's efficiency in processing visual information is compromised when faced with decreased visual clarity. This increased load on visual processing in blurred conditions underscores the importance of optimizing visual clarity in VR, where rapid and efficient visual information processing is critical.

\subsection{Limitation and Future Work}
One limitation in conducting the experiments is the potential weakness in evoking the P3 component. To enhance the distinct formation of the P3 component, the oddball paradigm is recommended \cite{kuziek2019real, kober2012using}, alongside increasing the number of subjects to boost event-related signal components. This approach leads to higher amplitudes of the P3 wave, allowing for a more precise examination of anomalies in stress stimuli.

Another constraint encountered was the utilization of a commercial EEG for investigating visual discomfort factors in a laboratory setting. This was particularly challenging for frontal electrodes due to points of contact between the Emotiv EPOC+ and the HMD. To obtain cleaner averaged ERPs, the EEG data analysis was restricted to electrodes of the parietal and occipital lobes, which were regions relevant to our analysis. Furthermore, commercial-grade EEG sensing devices may be integrated into VR headsets. Yet, in future work, we will compare the efficiency of commercial- and medical-grade devices regarding their ERP distinction accuracy. \added{Furthermore, ERPs can be affected by persons with neurodiversity or cognitive impairments \cite{le2024neurophysiological}. Hence, future work should consider persons with neurodiversity as participant samples.}

\added{Third, while we used object blurring and visual asymmetry to simulate visual discomfort associated with VAC, it is important to note that blurred vision is just one symptom of VAC, and our method does not replicate all the complex dynamics of VAC. Our approach was chosen for its simplicity and reproducibility in controlled laboratory settings, following some precedents in the field \cite{kooi2004visual}. However, this simulation does not fully encompass the multifaceted nature of VAC as observed in more diverse experimental setups. Future experimental designs will include independent variables such as depth cues \cite{hussain2023improving}, convergence demands \cite{elias2019virtual}, and ambient occlusion to simulate real-world VAC conditions more closely. Enhanced depth cues can intensify the VAC by manipulating perceived object distances, while dynamic vergence adjustments will test the visual system’s response to rapid changes in eye convergence \cite{hua2017enabling}. Additionally, ambient occlusion, which affects how light and shadows are rendered in 3D environments, will be manipulated to examine its effects on perceived spatial relationships and depth perception \cite{weier2018foveated, weier2017perception}, further enhancing the realism of VAC simulations. These methods aim to provide a more comprehensive understanding of VAC’s impact on visual discomfort in virtual environments.}

\added{Lastly, the correlation between ERP peaks and subjective measures of discomfort, such as the SSQ and visual discomfort questionnaire, was not directly analyzed due to the timing of these assessments. The SSQ and visual discomfort questionnaires were administered pre and post-task, preventing their coregistration with the EEG data captured during the task. This methodological setup was intended to facilitate an uninterrupted online evaluation of visual discomfort during the task. Future studies are proposed to incorporate continuous subjective reporting alongside real-time ERP monitoring to address this limitation. This enhancement will enable the direct correlation of ERP responses with subjective experiences of discomfort and symptoms reported through the SSQ.}

\section{Conclusion}
This research examined whether ERP can effectively detect visual discomfort. Our findings demonstrated a significant effect in the N2 component when participants were exposed to varying blur levels. However, the P1 and P3 components did not exhibit significant changes. Given that visual discomfort can arise from improperly adjusted interpupillary distances and lenses, we view our findings as an additional measure to assess participants' perceived experiences in studies utilizing ERPs as an objective gauge. Designers focused on user experience can use our findings to track the real-time visual discomfort of displayed elements. Additionally, our research shows the potential for adjusting head-mounted display settings in real-time to mitigate perceived visual discomfort in future VR applications.

\section{Open Science}
We encourage readers to review, reproduce, and extend our results and analysis methods. We make our collected datasets and Unity project available at this link {\url{https://osf.io/yvrw5/}.


\acknowledgments{
Francesco Chiossi was supported by the Deutsche Forschungsgemeinschaft (DFG, German Research Foundation), Project-ID 251654672-TRR 161. This work has been funded by the European Union’s Horizon 2020 research and innovation program under grant agreement No. 952026 (\url{https://www.humane-ai.eu/}). Thomas Kosch was supported by the German Research Foundation (DFG),
CRC 1404: “FONDA: Foundations of Workflows for Large-Scale Scientific Data Analysis” (Project-ID 414984028).
}

\bibliographystyle{abbrv}
\bibliography{main}
\end{document}